\documentclass[letterpaper]{article} %
\usepackage[preprint]{aaai24}  %
\usepackage{times}  %
\usepackage{helvet}  %
\usepackage{courier}  %
\usepackage[hyphens]{url}  %
\usepackage{graphicx} %
\usepackage{natbib}  %
\usepackage{caption} %
\usepackage{algorithm}
\usepackage{algorithmic}
\usepackage{subcaption}
\usepackage{booktabs}
\usepackage{multirow}

\usepackage{subcaption}
\usepackage{amsmath}
\usepackage{newfloat}
\usepackage{listings}
\DeclareCaptionStyle{ruled}{labelfont=normalfont,labelsep=colon,strut=off} %
\floatstyle{ruled}
\newfloat{listing}{tb}{lst}{}
\floatname{listing}{Listing}
\newcommand{\primarykeywords}{
(2) Learning;
(7) Multi-Agent Planning;
}

\usepackage[switch, modulo]{lineno}

\title{No Panacea in Planning: \\Algorithm Selection for Suboptimal Multi-Agent Path Finding}

\author {
    Weizhe Chen\textsuperscript{\rm 1}\equalcontrib,
    Zhihan Wang\textsuperscript{\rm 2}\equalcontrib,
    Jiaoyang Li\textsuperscript{\rm 3},
    Sven Koenig\textsuperscript{\rm 1},
    Bistra Dilkina\textsuperscript{\rm 1}
}
\affiliations {
    \textsuperscript{\rm 1}University of Southern California\\
    \textsuperscript{\rm 2}The University of Texas at Austin\\
    \textsuperscript{\rm 3}Carnegie Mellon University\\
    weizhech@usc.edu, zhihan.wang@utexas.edu, jiaoyangli@cmu.edu, skoenig@usc.edu, dilkina@usc.edu
}

\usepackage{bibentry}

\begin{document}

\maketitle

\begin{abstract}
Since more and more algorithms are proposed for multi-agent path finding (MAPF) and each of them has its strengths, choosing the correct one for a specific scenario that fulfills some specified requirements is an important task. Previous research in algorithm selection for MAPF built a standard workflow and showed that machine learning can help. In this paper, we study general solvers for MAPF, which further include suboptimal algorithms. We propose different groups of optimization objectives and learning tasks to handle the new tradeoff between runtime and solution quality. We conduct extensive experiments to show that the same loss can not be used for different groups of optimization objectives, and that standard computer vision models are no worse than customized architecture. We also provide insightful discussions on how feature-sensitive pre-processing is needed for learning for MAPF, and how different learning metrics are correlated to different learning tasks. 
\end{abstract}

\section{Introduction}

Multi-agent path finding (MAPF) is the problem of generating a set of collision-free paths for a team of agents with given the start and goal locations of each agent while minimizing some optimization objectives (such as travel times, travel distances, etc.) \cite{DBLP:conf/socs/SternSFK0WLA0KB19}. In recent years, MAPF has been an emerging research domain that has gotten a lot of attention because of its wide applications in warehouses, airport schedules, and autonomous driving. While the problem is proved to be NP-hard in certain cases~\cite{surynek2010optimization,YuAAAI13,MaAAAI16,NebelICAPS20} and also hard in practice in other cases, many algorithms are created to find optimal or suboptimal solutions. For the same standard benchmark, new algorithms can solve the scenarios faster and faster or give better solutions within the same time limit. However, there is no silver bullet in the MAPF research community: Even the latest algorithm does not perform the best in every scenario \cite{okumura2023improving, okumura2023lacam}, and different hyperparameters are needed for the best performance in different scenarios. 
Due to the extremely large number of possible combinations between algorithms and hyperparameters, running them all one by one or in parallel is impractical.
Therefore, for real-world deployment, when there is a previously known map with a strict time limit and solution quality requirement, choosing the specific algorithm and specific hyperparameters is a very important task, and algorithm selection in MAPF has become a focus in both academia and industry.

Researchers have already shown that machine-learning-based algorithm selection can learn to efficiently propose the fastest algorithm in specific scenarios \cite{DBLP:conf/aips/KaduriBS20}. 
Recent works has become proposing customized neural network architectures for the problem, and they want to show that this could help.
However, we believe these previous works have improperly built the workflow, and may lead to a wrong conclusion. 

In this paper, we address the algorithm-selection problem for solvers for MAPF, without the specific limit for focusing on the runtime of optimal algorithms only and without the specification that is selecting algorithms between completely different algorithms but also selecting between different hyperparameters for a single algorithm. Reducing these limitations give us broader potential applications, and enables us to show how the previous workflow could be misleading even in slightly different domains. We still formulate the problem as a prediction problem and use image-based machine learning to make the prediction. We propose multiple groups of objectives where each objective is a way to trade off the cost and the runtime. 
We realize that while previous researchers normally consider the optimization objectives, metrics for learning, and learning tasks in a pairwise independent manner, there are correlations between the three perspectives.
So, unlike other works, we propose a group of different learning tasks for different metrics, instead of using the same training scheme and evaluating it on multiple tasks.
Based on a newly constructed dataset built on the standard MAPF benchmark \cite{DBLP:conf/socs/SternSFK0WLA0KB19}, we use extensive experiments to show that the previous ways of using either interpolation or padding in all features are not correct in learning for MAPF, and using the same loss function all the time, which is what the prior works on algorithm selection for optimal MAPF did, can make the model perform poorly for specific objectives in algorithm selection for suboptimal MAPF. We also show how different machine learning models can make a difference in the algorithm selection performance, in which customized models show no advantage over standard computer vision models like ResNet and ViT. 
We also discuss how to choose the neural network used in the algorithm selection problem for MAPF that is fast, easy to train, and performs well.

\section{Related Work}

While there are a few works that directly generate solutions from machine learning \cite{Laurent21}, MAPF is nowadays mainly solved with more classic methods like heuristic search algorithms \cite{DBLP:journals/ai/SharonSFS15}, rule-based algorithms \cite{DBLP:journals/ral/HanY20}, and reduction-based algorithms \cite{DBLP:conf/ecai/SurynekFSB16}. There are mainly two groups of algorithms, namely optimal and suboptimal algorithms. The optimal algorithms like Conflict-Based Search (CBS) \cite{DBLP:journals/ai/SharonSFS15} are guaranteed to generate a solution that is optimal but usually requires a long runtime. Suboptimal algorithms usually generate good-quality solutions faster, but 
do not guarantee the solution found to be the best one. 
Within suboptimal algorithms, the algorithms can be further divided into two groups, namely, ones that still guarantee the solution quality to be within a suboptimality bound, which is also known as asymptotically optimal algorithms, like Explicit Estimation CBS (EECBS) \cite{LiIJCAI21}, and ones without any guarantees like Prioritized Planning (PP) \cite{DBLP:conf/aiide/Silver05}, Parallel Push-and-Swap (PPS) \cite{DBLP:conf/socs/SajidLB12}, and 
Priority Inheritance with Backtracking (PIBT+) \cite{okumura2022priority}. 
In this paper, we consider optimal algorithms and both types of suboptimal algorithms as candidate algorithms.

Algorithm selection is the problem of selecting a specific algorithm for a specific scenario \cite{Smith-Miles08}. It has been successfully used in many optimization problems, including satisfiability and traveling salesman problem (TSP)  \cite{DBLP:journals/ec/KerschkeKBHT18,DBLP:conf/sat/XuHHL12}. The rich studies in those domains have built up a standard way of measuring the performance of selected algorithms for their specific usage. However, algorithm selection in MAPF is still at an early stage, where most papers \cite{DBLP:journals/corr/abs-1906-03992,DBLP:conf/atal/RenSESA21} just show that,  with a specific technique, the selection algorithm could be helpful in choosing a correct solver algorithm but lacked thorough performance metrics. %
\citet{DBLP:journals/corr/abs-1906-03992} is the first to introduce the algorithm selection problem to MAPF. They proposed a modified version of AlexNet \cite{DBLP:conf/nips/KrizhevskySH12} and demonstrated its ability to predict the fastest algorithm in MAPF. \citet{DBLP:conf/aips/KaduriBS20} improved the results by using VGGNet \cite{DBLP:journals/corr/SimonyanZ14a} and gradient boosted decision tree with XGBoost \cite{DBLP:conf/kdd/ChenG16}. \citet{DBLP:conf/atal/RenSESA21} proposed MAPFAST, which added more features related to the shortest path between the start and goal locations of each agent. They also added two auxiliary output channels into the loss function, and used an inception-based neural network \cite{DBLP:conf/cvpr/SzegedyLJSRAEVR15} to train the model. 
Recently, MAPFASTER\cite{DBLP:conf/iros/AlkazziRSM22} is proposed to improve the training and inference speed of the neural network and try to fix a problem for rescaling the inputs that occur in MAPFAST\cite{DBLP:conf/atal/RenSESA21}.
\citet{DBLP:conf/socs/KaduriBS21} empirically showed that different types of maps have different preferences for different algorithms, and thus confirmed the usefulness of algorithm selection. Previous works focused on either only the accuracy of finding the fastest or only the coverage rate, which is how likely the chosen algorithm can finish in a given time limit. Since both criteria are only related to the runtime of the algorithms, in this paper, we additionally focus on suboptimal algorithms, and consider how to take both runtime and cost into account simultaneously. We also explore modern neural network architectures to validate if modern architectures can help in prediction.

In this paper, we formulate the problem as an image-based prediction problem. Many works can be used to solve the problem. 
AlexNet \cite{DBLP:conf/nips/KrizhevskySH12} was the first model showing that deep learning can succeed in class prediction. 
VGGNet \cite{DBLP:journals/corr/SimonyanZ14a} was then proposed to show that using a small kernel size can stably predict good results while keeping the total number of training parameters low. 
GoogLeNet \cite{DBLP:conf/cvpr/SzegedyLJSRAEVR15} was proposed about the same time for the same purpose but proposed an inception unit as their solution. 
ResNet \cite{DBLP:conf/cvpr/HeZRS16}  proposed to use skip links between layers to solve the vanishing gradient problem, making very deep architectures possible, and has become one of the most commonly used models in computer vision. 
In recent years, Transformer \cite{DBLP:conf/nips/VaswaniSPUJGKP17} has become increasingly popular in all machine learning research, including computer vision. Vision Transformer (ViT) \cite{DBLP:conf/iclr/DosovitskiyB0WZ21} proposed to split the image into smaller blocks for transformer to use, and is now a popular and successful model in computer vision. 
In this paper, we test the aforementioned neural networks that are popular in computer vision and algorithm selection in MAPF research.

\section{Preliminaries}
The multi-agent path finding (MAPF) problem is the problem of finding a set of conflict-free paths for a set of agents in a known environment while minimizing their travel times. Specifically, in this paper, we consider exactly the same problem as \cite{DBLP:conf/socs/SternSFK0WLA0KB19,DBLP:conf/aaai/0001CHSK22}, which is a four-connected grid map, where each agent is given a start cell and a goal cell. A scenario is defined as the combination of the description of the map, which is the size of the map and which cells have obstacles, the start cells, and the goal cells of each agent. At each timestep, an agent can move to an adjacent cell or stay in its current cell. A conflict happens if two agents result in the same cell at the same timestep. Each agent remains at its goal cell after it arrives until every agent arrives at their goals. The quality of one solution is the total sum of travel times of each agent, which is also known as the sum of costs in the MAPF community.

Algorithm selection is the problem of choosing a suitable algorithm that is the best in a given scenario.
Generally, algorithm selection includes both selecting completely different algorithms and choosing the hyperparameter(s) of a fixed algorithm. (Our setting of hyperparameter selection is different from algorithm configuration \cite{DBLP:journals/jair/EggenspergerLH19} in that we are given a small set of candidate hyperparameter combinations ahead, and we want to choose the best one for a new scenario without any trails.) For clarity, we use the word solver to refer to these different candidates. However, we still call the problem algorithm selection to be consistent with other papers that solve the same problem. The input for the algorithm selection includes precisely the same information as a MAPF solver knows without adding any information from each candidate solver. 
The selection algorithm is then required to output the best solver.
Typically, machine-learning-based selection algorithms first transform the scenario information to desired formats and features they would like to use as input and then use the learning model to output the answer directly. We follow the same workflow in this paper.

In this paper, we formulate our algorithm selection as a standard image-based-input prediction problem solved by computer vision techniques. 
It is remarkable that, in the last decade, computer vision has earned a great improvement with tens of thousands of papers every year. There are many new neural network models created and evaluated on standard datasets like ImageNet \cite{yang2019fairer} and CIFAR-10 \cite{krizhevsky2009learning}, and many models have successfully been used in different applications like autonomous driving, face recognition, and many novel applications. These successful applications, in turn, encourage the investment and development of computer vision. 
Furthermore, these models have successfully shown that with specific data augmentation, raw image input without human-involved complex pre-processing can be as good as using more manually designed features. 
That is why we would like to explore many modern computer vision models to see if they can directly generate some good results. 
However, as we will later show in the experiment section, we find that there is no panacea to planning: Using the same model with the same loss function does not work for all objectives.

\section{Algorithm Selection for Suboptimal MAPF}
\subsection{Dataset}

Because there is no standard dataset for algorithm selection in MAPF right now, %
we first describe how we build our own datasets before we describe our learning tasks. 

\subsubsection{Candidate Solvers}

In this paper, we build two separate datasets for the purpose of doing standard algorithm selection between different algorithms, and for the purpose of doing solver selection between different solvers that are from the same algorithm.

For the standard algorithm selection part, we enumerate some of the typical MAPF solvers nowadays. Our candidate algorithms are: CBS \cite{DBLP:journals/ai/SharonSFS15}, EECBS \cite{LiIJCAI21}, PP \cite{DBLP:conf/aiide/Silver05}, PPS \cite{DBLP:conf/socs/SajidLB12}, and PIBT+ \cite{okumura2022priority} . We did not include some recent research, such as Large Neighborhood Search (LNS) \cite{LiIJCAI21}, which can use various MAPF solvers as subsolvers. LNS-based solvers can solve any scenarios that are solvable by other solvers by using them as the initial solver, with just a small overhead, and can solve additional scenarios when the later neighborhood search is performed. Therefore, comparing LNS-based solvers to more standard MAPF solvers is generally not fair. However, theoretically, LNS users can also use our model to choose which algorithm they want to use as their subsolvers.

For the hyperparameter selection, we want to select the optimality bound for EECBS. The optimality gap $w$ in EECBS guarantees the solution found to be within $w$ times compared to an optimal solution, and more time will be spent during the search almost for sure if a small and tight bound is given. But a large $w$ can also find near-optimal or even optimal solutions in specific scenarios, so choosing the optimality gap can make the runtime much faster in those scenarios where giving a small value is not helpful for EECBS to find a better solution. Specifically, we aim to choose the optimality gap from $w={1.05, 1.1, 1.15, 1.2}$.

\subsubsection{Features} \label{sec:features}
Features have long been known as a very important factor in machine learning research related to MAPF. Many early works on algorithm selection for MAPF used the scenario information without any pre-processing \cite{DBLP:journals/corr/abs-1906-03992,DBLP:conf/aips/KaduriBS20}. One exception \cite{DBLP:conf/atal/RenSESA21} considered only a given shortest path without specifying a clear way to determine which shortest path to use when an agent has multiple shortest paths. It also over-compressed different kinds of information into each channel, which can largely reduce the potential performance. So, in this work, 
we use the image representation of the MAPF scenarios from the MAPF benchmark and use a set of features, each one of which is a separate channel fed into the model as input. Given the fact that feature usefulness validation is very complex for neural-network-based approaches, and is not the main focus of this work, here we just list the features used in this paper in order: 
\begin{enumerate}
	\item Whether this specific cell is an obstacle. This is always the same in scenarios generated from the same map.
	\item Whether this specific cell is a start cell. Different start cells do not have any further differences.
	\item Similar to start cells, an indicator of whether the specific cell is a goal cell for any agent.
	\item If everyone follows its shortest path selected by \cite{DBLP:journals/ral/HanY20} without considering any conflict, how many times will the cell be visited.
	\item Total number of conflicts on the cell between all pairs of shortest paths from different agents. 
	\item Total number of conflicts on the cell between all pairs of shortest paths and 1-suboptimal paths from different agents that happens on the cell. 
	\item How many times will the cell be visited if every agent tries every possible shortest path on its own without considering any conflict.
\end{enumerate}
Here, we define a 1-suboptimal path as a path whose length is the length of the shortest path plus 1. 
In this paper, we make the input a rectangle image, and 
then we normalize all the input features to a scale from 0 to 100 by dividing every channel with the maximum value of all maps for a specific channel and multiplying the value with 100.

During our experiments, we realize that although cells with obstacles can never be occupied or visited by any agents, treating those cells with a value of 0 in all channels is not always helping the learning. 
Some features provide information about the level of congestion in the scenario. Therefore, the scenario should be considered more crowded and have a larger value on the features when there are more obstacles on the map, and less crowded when fewer obstacles are present.
So, we change all obstacle cells to a value of 200 for the (4) and (7) channels listed above, which is the heatmap that sums up the shortest path from a given set, based on some empirical results on a not complete enumeration and the intuition we mentioned above.

After defining the features, we need to pre-process the data to make the input fit into the same neural network so we do not need to train separate models for different maps. While previous works \cite{DBLP:journals/corr/abs-1906-03992} all use a default resize\footnote{Researchers in MAPFAST \cite{DBLP:conf/atal/RenSESA21} said they use padding to formalize the input but in fact, they are not using it according to their public code on Github.}, which is also known as interpolation in many fields, to make all image to be of the same size without even mentioning that in the paper, we realize that this is not always the correct thing to do. 
This resizing way works properly in computer vision because the pictures have a good property of zooming invariance, i.e., a hand is still a hand no matter how it is zoomed in or zoomed out. However, in MAPF, each agent can only move one cell at one timestep, so zooming invariance is not held. Therefore, using resize as the way to rescale images of different sizes to the same sizes is not a proper way. We give another option for pre-processing the data, which is for each given image, we make the original image in the center, and pad the image to a fixed size of $384 \times 384$ with the number we get from an obstacle. Specifically, the value padded is not a fixed number in different channels, because the padded part should be equivalent to padding obstacles to the map and obstacles are not necessarily getting the same value in different channels. 

\subsubsection{Labels}

With all the candidate solvers, we run all solvers on all maps in the MAPF benchmark \cite{DBLP:conf/socs/SternSFK0WLA0KB19}. For each solver, we enumerate different numbers of agents with a step size of 10 agents until there are no more than 2 solvers that can solve any larger scenario within the time limit of 2 minutes on the map. With all these statistics together with the features we have just defined, we get an algorithm selection dataset of 89,940 data points and a hyperparameter selection dataset of 53,691 data points. It is noteworthy that the algorithm selection dataset is much larger compared to any previously used dataset that is around 10K \cite{DBLP:conf/aips/KaduriBS20,DBLP:conf/atal/RenSESA21}, and a reasonable size for a start trying modern deep learning models that have millions of parameters without immediately overfitting, but still quite small compared to standard datasets in machine learning that normally have millions of datapoints. 

\subsection{Optimization Objectives}

\begin{figure}[t]
	\centering
		\includegraphics[width=0.98\linewidth]{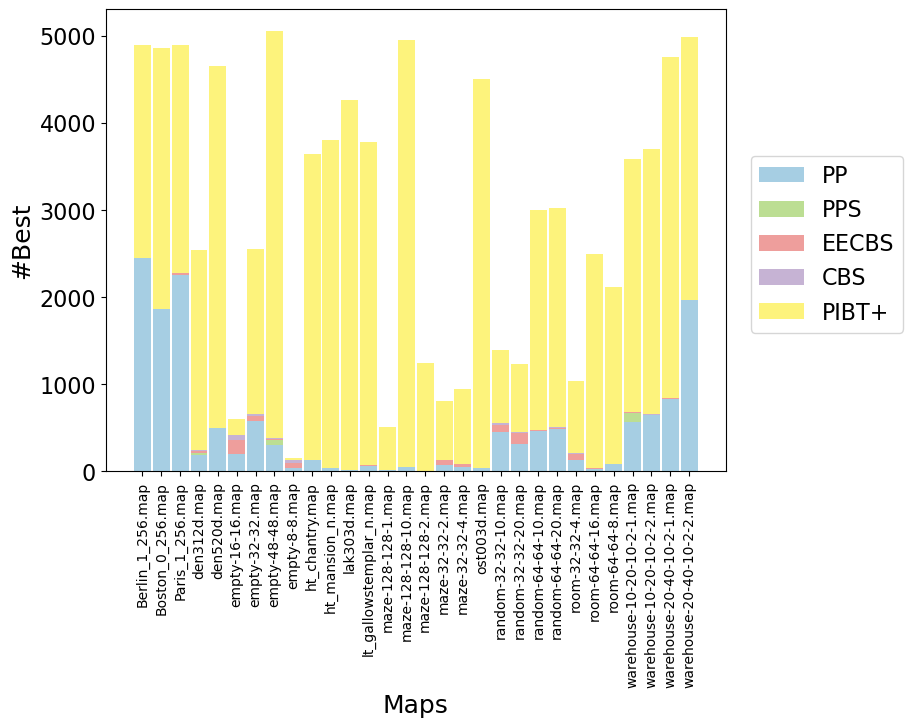}
		\caption{The frequency of each algorithm being the best one on a given map when $w=0.001$ in Eq.~\ref{eq:score_def}. Different maps have different numbers of total scenarios because different maps are different in size and obstacles so the total capacity is naturally different. }
		\label{fig:optimal_freq}
\end{figure}

Previous works on algorithm selection for MAPF typically consider the runtime or success rate (at a fixed runtime threshold) of a solver as the only objective and never take solution quality into account. In this way, they can hopefully always deliver a solution to the deployed scenario. However, the actual solution quality can vary largely across different solvers in different maps when we consider suboptimal solvers. So we need to consider both runtime and cost (solution quality) simultaneously. While there are many ways to do this, we use this section to give two intuitive ways of defining the optimization objective. 

We first normalize the time and the cost to a similar scale, in a way that is independent of the set of candidate solvers. Suppose the time limit is $time_{limit}$, and the sum of costs for shortest paths for every single agent is $cost_{bound}$ in the scenario, the normalized time and cost are calculated by:
\begin{align}
	time' &= \frac{time}{time_{limit}} \\
	cost' &= \frac{cost}{cost_{bound}},
\end{align}%
Furthermore, following the previous convention, if a specific solver cannot be finished in the time limit $time_{limit}$ in any scenario, the time used for calculation is defined as $5 \times time_{limit}$, and the cost used for calculation is $5 \times cost_{min}$, where $cost_{min}$ are the minimum sum of costs in all success candidate solvers (i.e., all candidate solvers that find solutions within the time limit). For clarification, the number 5 used here is an example number used in this paper that is big enough to impose a penalty for those who time out, but also not too large to draw excessive attention to this specific data point. Other numbers, such as 2 or 10, will also likely work.

The first and the most common way of defining an objective is to use a weighted sum of different metrics as the score \cite{DBLP:journals/ai/BischlKKLMFHHLT16,DBLP:conf/foga/HeinsBPSTK21,DBLP:conf/ppsn/SeilerPBKT20}. 
Specifically, we calculate the following objective $score$:

\begin{align}
	score(a) = time'_a + w * cost'_a, \label{eq:score_def}
\end{align}%
where $w$ is the hyperparameter that users can control to represent their preference, and $time'$ and $cost'$ are the normalized metrics we calculated above. In this case, we want to find the solver with the best score:

\begin{align}
	\min_a & \quad score(a) \label{eq:score_task}
\end{align}

The second group of objectives is to choose a solver that gives a solution within a given cost bound the fastest. Specifically, we find the solver $a$ so that:
\begin{align}
	\min_a & \quad time_a \nonumber\\
	s.t. & \quad cost_a \le bound * cost_{min}, \label{eq:bound_task}
\end{align}%
where $cost_{min}$ is the minimum sum of costs in all success candidate solvers. This group of objectives is useful in the case that the users just need some guarantee on the solution quality, but as long as the solution is a relatively good one, the runtime becomes the only consideration.

With different objectives, we have a few data points with the same input but different labels from one scenario. In Fig.~\ref{fig:optimal_freq}, we show the frequency of different solvers being the best in different scenarios, and we further provide the frequency plots of all unique tasks in the appendix. We observe that the best solvers are changing a lot while trading off the runtime and cost at different weights.
And it is also different between maps in how large the difference between single best solvers and other solvers is even if the single best solvers are the same.

\subsection{Metrics for Learning}

\begin{figure}[t]
	\centering
	\includegraphics[width=0.5\linewidth]{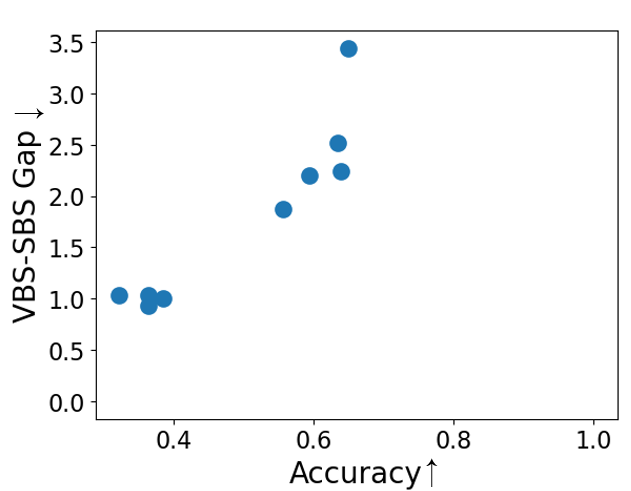}
	\caption{Overall performance of different algorithm selection models in different tasks in terms of accuracy (higher is better) and VBS-SBS gap (lower is better). Blue points are actual samples we collected in different models.}
	\label{fig:acc_gap}
\end{figure}

After defining the optimization objectives, we briefly talk about the two metrics for learning we use in this paper. 

The first metric is the standard accuracy, which is used both in popular machine learning research, and in earlier papers in algorithm selection for MAPF \cite{DBLP:conf/atal/RenSESA21}.

Accuracy is normally a good metric in ML, but it only takes right or wrong into account, not how wrong the predictions are. On the other hand, if we want to use an algorithm selector, we definitely want to know how bad the wrong choices are as long as the selector is not perfect. So the second group of objectives is a well-known metric in algorithm selection, the VBS-SBS gap \cite{DBLP:conf/sat/XuHHL12}, which is also known as the closed gap \cite{DBLP:conf/oasc/LindauerRK17}. 
(The VBS-SBS gap equals $1$ minus the closed gap.)
Throughout the text, we will use the term "gap" to refer to the VBS-SBS gap. 
Given a pre-calculated virtual best solver (VBS) that always correctly outputs the best solver in each scenario and a single best solver (SBS) that is always outputting the same posterior best solver, the gap for the current solver $a$ is calculated by:

\begin{align}
	gap(a)=\frac{\overline{score(a)}-\overline{score(VBS)}}{\overline{score(SBS)} - \overline{score(VBS)}},
	\label{eq:vbs-sbs-score}
\end{align}%
where $\overline{score}$ is the average score function over all data points, which could be the score function we defined in Eq.~\ref{eq:score_def}, or just runtime if we use the second optimization objective in Eq.~\ref{eq:bound_task}. This objective is commonly known as a metric for a prediction model, with 0 as the theoretical best and 1 as the same performance as the prediction model that always outputs the single best solver. A good algorithm selector should have a gap smaller than 1, and the smaller the gap is, the better the selector is. 

However, on some optimization objectives, the gap metric does not get better in the same direction with accuracy. 
When a user prioritizes accuracy optimization, the model will likely converge to weights that favor common scenarios. Although this approach may seem reasonable, it can produce the least favorable results if the common scenarios are incorrect. Consequently, this can lead to poor performance on metrics that penalize incorrect answers, and the VBS-SBS gap is one of these metrics.
As shown in Fig.~\ref{fig:acc_gap}, having good accuracy does not necessarily guarantee a small VBS-SBS gap, and, in most cases, achieving a small gap requires a drop in accuracy.
Therefore, we believe that the previous workflow of training a model and then evaluating it with both metrics is not the appropriate approach here in MAPF, and users should build different learning tasks after knowing what they truly want.

\subsection{Learning Tasks}

In this paper, based on the optimization objectives we have, we consider two popular ways of learning an algorithm selection model:

\begin{enumerate}
	\item{Classification}: The standard way is to treat the problem as a classification problem. The model predicts the probability of choosing each solver and is trained with typical classification loss. For inference, the solver with the largest probability is selected.
	\item{Regression on Expected Score}: This method still predicts the probability of each candidate being the best, but it now uses a regression-based loss instead of a classification-based loss during training. Every time, the probability got from the output is used to calculate the expected score by using the model following the probability, and the loss is a function related to how much difference this score is compared to the VBS.
\end{enumerate}

Specifically, we choose to use two variants of classification and one way of regression.

The first variant of classification is the most popular version, which is to use a cross-entropy loss for the classification problem. We call this method CE (cross-entropy).

The second variant of classification is by simply changing the cross-entropy loss to a binary cross-entropy loss,
which we call BCE (binary cross-entropy).

The third way of learning is the regression on the expected scores we described above. With the probability obtained from the model, we compute Huber loss \cite{huber1992robust} for our expected score and the score of VBS as the loss. We call this learning way Reg (Regression).

A learning \textit{task} is the combination of optimization objectives and a specific learning way.
We use Score-w-y to refer to the optimization task in Eq.~\ref{eq:score_task}, where w is $w$ from Eq.~\ref{eq:score_task} and y is the learning way to create a learning task. We use Bound-b-y to refer to tasks optimizing as Eq.~\ref{eq:bound_task}, where b is the value of bound in the equation.
For example, Score-0.001-CE is the task that we set $w=0.001$ in Eq.~\ref{eq:score_def} for score definition, and we use cross entropy in the training. 
Bound-1.1-BCE means the optimization tasks with $bound=1.1$ in Eq.~\ref{eq:bound_task}, and optimized by the binary cross-entropy loss.

\section{Experiments}

\subsection{Experiment Settings}

In this paper, we develop our machine learning model based on 5 different computer vision models: VGG16 (VGGNet) \cite{DBLP:journals/corr/SimonyanZ14a}, ViT-Tiny (ViT) \cite{DBLP:conf/iclr/DosovitskiyB0WZ21,rw2019timm}, MAPFAST \cite{DBLP:conf/atal/RenSESA21}, MAPFASTER \cite{DBLP:conf/iros/AlkazziRSM22}, ResNet-18 \cite{DBLP:conf/cvpr/HeZRS16}, based on the standard Timm library \cite{rw2019timm}. These models have covered all existing literature in deep-learning-based algorithm selection for MAPF and also the most commonly used models in computer vision. Surprisingly, we found that the auxiliary tasks used in MAPFAST are not helping the result for any tasks in our dataset, so all numbers of MAPFAST are from models trained with only the first output channel that directly outputs the probability of each solver, and optimized as a standard classification problem. 
We do not include any results based on decision trees because previous papers in algorithm selection for MAPF \cite{DBLP:conf/aips/KaduriBS20, DBLP:conf/atal/RenSESA21} have already shown that neural networks can outperform decision tree models in most cases, and also decision tree models require a much more complex study of which features to use, which is not the focus for this paper.
It needs to be addressed that ViT is very different from any other model because it is not a convolutional neural network. We hope that ViT can perform differently because it does not have the limit of the small kernel size, nor the shifted invariance property that CNN generally has.

For each task with each model, we show the accuracy and VBS-SBS Gap  (Gap) for them. We choose to show both of them because accuracy is the currently common practice for algorithm selection for MAPF, while the Gap is another metric that is widely used in more general algorithm selection. While we primarily set our baseline as getting a better number than SBS, our SBS is selected separately for accuracy ($\text{SBS}_{\text{Acc}}$) and VBS-SBS gap ($\text{SBS}_{\text{Gap}}$). From intuition, $\text{SBS}_{\text{Acc}}$ is the solver that is the most common in Fig.~\ref{fig:optimal_freq}, while $\text{SBS}_{\text{Gap}}$ is the one that can solve the most instance, therefore it does not get any large failing penalty. Because of the dominance of each solver in our setting, the SBS for the entire dataset is the same as the SBS per grid or per map type.

For every model, we use some data augmentation methods to prevent the model from overfitting, which include random flip, random rotation, and random erasing with a probability of $0.5$. We decide to use resize only in the 4th to the 6th channels of the feature and padding in other channels because they are empirically the best as we will later show.

The running results of all MAPF solvers are collected on the same AWS EC2 m4 server, while the training and testing of all machine learning models are conducted on a xeon-6130 server with a single NVIDIA-V100 and 184GB RAM. All parameters are selected by grid-search, and the full hyperparameter table is provided in the appendix.

\begin{table}[t]
	\centering
	\resizebox{0.75\columnwidth}{!}{
		\begin{tabular}{l|l|l}
			Padding(\texttt{p}), Resize(\texttt{r}) & Gap $\downarrow$   & Acc $\uparrow$  \\ \hline
			\texttt{ppp pppp}              & 1.000 & 0.262 \\
			\texttt{ppp prrp}              & 1.100 & 0.298 \\
			\texttt{ppp rppp}              & 1.000 & 0.262 \\
			\texttt{ppp rrrp}              & \textbf{0.911} & 0.307 \\
			\texttt{ppp rrrr}              & 1.265 & 0.305 \\
			\texttt{rrr rrrp}              & 1.118 & 0.297 \\
			\texttt{rrr rrrr}              & 0.945 & 0.279
	\end{tabular}}
	\caption{Partial results on different rescale methods on different features trained with ViT on Score-1-Reg task. 'p' denotes using padding and 'r' denotes using interpolation (default resize) in the corresponding channel of features, in the order shown in the \textit{Features} section. The full table can be found in the appendix.}
	\label{tab:resize}
\end{table}

\subsection{Feature Rescale} \label{sec:rescale}
In earlier papers using image-based models \cite{DBLP:conf/aips/KaduriBS20,DBLP:journals/corr/abs-1906-03992,DBLP:conf/atal/RenSESA21,DBLP:conf/iros/AlkazziRSM22}, researchers are always using interpolation resize (known as resize in ML libraries like Pytorch \cite{DBLP:conf/nips/PaszkeGMLBCKLGA19}) to make the inputs to a fixed size of $227 \times 227$. Recent algorithm selection in MAPF papers are following this tradition from computer vision. However, resizing the pictures by interpolating the values on every pixel from the original input is losing many underlying assumptions in planning. For example, we can only move one cell at a time, so a cell of $1\times 1$ in a $10\times 10$ map is very different from a group of cells of $10 \times 10$ in a $100 \times 100$ map. On the other hand, always using the padding to rescale features is not aligned with our intuition in heatmap features where we probably only care about the overall crowdness. Given that it is hard to collect a lot of data, we want to make full use of our data and features like heatmap should be used across different sizes of map. So in this section, we examine how the performance of a model will be if we change channels from interpolation to padding.

\begin{table*}[t]
	\centering
	\begin{tabular}{@{}llllllll@{}}
\toprule
         &                 & MAPFAST & ViT           & VGGNet        & ResNet        & MAPFASTER & SBS  \\ \midrule
         & Score-0.001-CE  & 0.81    & 0.81          & 0.84          & \textbf{0.91} & 0.8       & 0.8  \\
         & Score-0.001-BCE & 0.85    & 0.83          & 0.85          & \textbf{0.86} & 0.8       & 0.8  \\
         & Score-0.1-CE    & 0.61    & 0.61          & \textbf{0.69} & 0.61          & 0.57      & 0.57 \\
Standard & Score-0.1-BCE   & 0.65    & 0.62          & \textbf{0.67} & 0.56          & 0.57      & 0.57 \\
         & Score-1-CE      & 0.62    & \textbf{0.69} & 0.66          & 0.6           & 0.62      & 0.62 \\
         & Score-1-BCE     & 0.57    & \textbf{0.68} & 0.64          & 0.62          & 0.62      & 0.62 \\
         & Bound-1.1-CE    & 0.56    & 0.65          & \textbf{0.69} & 0.61          & 0.51      & 0.51 \\
         & Bound-1.2-CE    & 0.72    & \textbf{0.74} & 0.73          & 0.58          & 0.19      & 0.51 \\ \midrule
EECBS    & Score-0.001-CE  & 0.7     & \textbf{0.71} & 0.7           & 0.69          & 0.69      & 0.69 \\
         & Score-0.001-BCE & 0.69    & 0.69          & \textbf{0.7}  & 0.69          & 0.69      & 0.69 \\ \bottomrule
\end{tabular}

	\caption{The accuracy for different models in different tasks. The names of the tasks are shown in the Task column, following the naming described in the experiment setting section.}
	\label{tab:acc_table_model}
\end{table*}
\begin{table*}[t]
	\centering
	\begin{tabular}{@{}lllllll@{}}
\toprule
                       & MAPFAST       & ViT           & VGGNet & ResNet        & MAPFASTER & SBS \\ \midrule
Score-0.001-Reg         & 0.89          & \textbf{0.87} & 1      & 0.94          & 1         & 1   \\
Score-0.1-Reg           & 0.75          & \textbf{0.71} & 1      & 1             & 1         & 1   \\
Score-1-Reg             & \textbf{0.71} & 0.92          & 1      & 0.75          & 1         & 1   \\
Score-0.001-Reg (EECBS) & 1             & 1             & 1      & \textbf{0.93} & 1         & 1   \\ \bottomrule
\end{tabular}

	\caption{The VBS-SBS gap for different models in different tasks. The names of the tasks are shown in the Task column, following the naming described in the experiment setting section. The smaller the number, the better the model. The first three rows are from the standard dataset, while the last one is from our hyperparameter selection for EECBS dataset.}
	\label{tab:gap_table_model}
\end{table*}
\begin{table}[t]
	\centering
	\begin{tabular}{@{}lllll@{}}
\toprule
                  & CE            & BCE           & Reg           & SBS  \\ \midrule
Score-0.001-Acc$\uparrow$   & 0.81          & \textbf{0.83} & 0.79          & 0.8  \\
Score-0.001-Gap$\downarrow$   & 19.03         & 18.67         & \textbf{0.87} & 1    \\
Score-0.1-Acc $\uparrow$    & 0.61          & \textbf{0.62} & 0.57          & 0.57 \\
Score-0.1-Gap$\downarrow$     & 14.83         & 14.79         & \textbf{0.71} & 1    \\
Score-1-Acc $\uparrow$      & \textbf{0.69} & 0.68          & 0.3           & 0.62 \\
Score-1-Gap$\downarrow$       & 1.95          & 1.89          & \textbf{0.92} & 1    \\
Score-0.001-Acc(EECBS)$\uparrow$ & \textbf{0.71} & 0.69          & 0.69          & 0.69 \\
Score-0.001-Gap(EECBS)$\downarrow$   & 1.12          & \textbf{1}    & \textbf{1}    & 1    \\ \bottomrule
\end{tabular}

	\caption{The Accuracy and Gap of ViT trained with different loss in different dataset. We provide the results of all model architectures in the Appendix. The last two rows are results from the hyperparameter selection for EECBS dataset, while the other rows are from the standard dataset.}
	\label{table:metrics-loss}
\end{table}

Because enumerating all possible combinations of padding and resizing in different channels needs $2^7=128$ experiments, which is too large for us to test them all, we assume that channels that have similar meanings should be treated the same, and thus, we can change the rescaling method in groups, i.e., features (1), (2), (3) are in a group, features (5) and (6) are in a group, and features (4) and (7) each is in a group.
We show the results of different combinations of interpolation (resize) and padding in Table.~\ref{tab:resize} and the appendix for the full enumeration. 
We found that choosing to use which rescaling method can make a different, and change one channel to another is not independent of what other channels currently are. In our setting, we found that the best rescaling method is to use padding in the first three channels that describe the MAPF instance as start locations, goal locations and obstacles, and the last channel, which is the heatmap of the sum of all possible shortest paths. All other channels should use interpolation. While we do find this really helpful in our experiment, we encourage later user to double-check if the same holds in their dataset given that the advantage is not very significant in our experiment.

\subsection{Model Architecture}

We show our comparison of different model architectures in  Table~\ref{tab:acc_table_model} and Table~\ref{tab:gap_table_model}. We found that while previous researchers claim that model architectures they proposed can be better than earlier models, their architectures fails to surpass popular computer vision models like ResNet and ViT. And overall, they even fail to reach the same level of performances as the standard models. In our opinion, the popular models are designed to efficiently extract both high-level and low-level information from the given pictures, and thus can be usually better than customized models.

Between all the architectures, there is no only winner. And given that ViT wins half of the tasks in both accuracy and Gap, and that its underlying transformer architecture is shown to be generally robust in foundation models, we recommend ViT as the default choice. VGGNet is strong in accuracy, but fails to jump out of the local minimum in the Gap metrics. MAPFAST performs good in most cases, but it does not show significant advantage to popular models. 
Specifically, MAPFASTER \cite{DBLP:conf/iros/AlkazziRSM22} never succeeds in escaping the local optimal of outputting SBS, which we believe is because of the over-compressed neural network structure.

In all groups of optimization objectives, there are always some learning tasks with some machine learning models that get a gap smaller than 1, and better accuracy compared to SBS. This means that algorithm selection in suboptimal solvers and hyperparameter selection between different solvers from the same algorithm are both feasible. On the other hand, the performances of learning models are different from task to task, and part of them shows a marginal improvement compared to the SBS. This indicates a great potential for future researchers to develop new methods to better solve these hard tasks of selecting suboptimal solvers. For clarification, the results from different weights are not directly comparable since the distribution of each algorithm being the best is completely different under different weights, making the difficulty of each task also different.

\subsection{Metrics-Loss Correspondence}

As we discussed earlier, optimizing accuracy could be the opposite of optimizing gap. In Table~\ref{table:metrics-loss}, we show the performance of models trained in different losses evaluated under different loss. It shows that there is no learning task that can be good in both accuracy and VBS-gap. Regression is the most promising way to build a learning task when using the VBS-SBS metrics, while CE and BCE are better when using the accuracy metrics. This means that the classic way of always using CE to train a model for all metrics used in previous papers \cite{DBLP:conf/aips/KaduriBS20, DBLP:conf/atal/RenSESA21}, is not preferable when Gap is included as one of the metrics.

\section{Conclusion}

In this paper, we study algorithm selection for suboptimal MAPF solvers. 
We formulate algorithm selection as a prediction problem with multiple potential formulations to trade off between runtime and solution cost differently. Deep learning models are trained with many combinations of optimization objectives and loss functions to make predictions.  

We showed that different metrics in the domains are not aligned with each other, and models optimized for accuracy could even go to the opposite direction to the models optimized for VBS-SBS gap.
We show that feature-specific ways of rescaling is needed, and that customized neural network architectures do not generally outperform standard models from computer vision domains. 
We hope our findings can change some of the common practices for algorithm selection for MAPF that might not be correct.
In addition, we show that hyperparameter selection can be successfully done with the same framework.

\bibliography{aaai24}

\appendix
\appendix

\section{Rescaling}
Here we give the full table of the ablation study of the rescaling method in the main paper.
\begin{table}[h!]
\centering
\begin{tabular}{l|l|l}
Padding(\texttt{p}), Resize(\texttt{r}) & Gap   & Acc   \\ \hline
\texttt{ppp pppp}              & 1.000 & 0.262 \\
\texttt{ppp pppr}              & 1.000 & 0.262 \\
\texttt{ppp prrp}              & 1.100 & 0.298 \\
\texttt{ppp prrr}              & 1.286 & 0.303 \\
\texttt{ppp rppp}              & 1.000 & 0.262 \\
\texttt{ppp rppr}              & 0.940 & 0.280 \\
\texttt{ppp rrrp}              & \textbf{0.911} & 0.307 \\
\texttt{ppp rrrr}              & 1.265 & 0.305 \\
\texttt{rrr pppp}              & 1.000 & 0.262 \\
\texttt{rrr pppr}              & 0.995 & 0.262 \\
\texttt{rrr prrp}              & 0.959 & 0.274 \\
\texttt{rrr prrr}              & 0.923 & 0.290 \\
\texttt{rrr rppp}              & 1.008 & 0.262 \\
\texttt{rrr rppr}              & 0.988 & 0.262 \\
\texttt{rrr rrrp}              & 1.118 & 0.297 \\
\texttt{rrr rrrr}              & 0.945 & 0.279
\end{tabular}

\caption{Results on different rescale method on different features. 'p' denote using padding, while 'r' denote using the default resize in the corresponding channel of features. The table is generated with results trained with ViT on Score-1-Reg task.}
\end{table}

\section{Full Results of Different Models}
Here we provide the full table instead of the ViT only version provided in the main paper as a reference.

\begin{table*}
    \centering
    \resizebox{0.9\linewidth}{!}{

\begin{tabular}{c|c|cc|cc|cc|cc|cc|c|c}
 \multicolumn{1}{l|}{}& \multicolumn{1}{l|}{}                  & \multicolumn{2}{c|}{$\text{MAPFAST}_{\text{cl}}$}                                & \multicolumn{2}{c|}{ViT}                                                      & \multicolumn{2}{c|}{VGGNet}                                                   & \multicolumn{2}{c|}{ResNet}                                                 & \multicolumn{2}{c|}{MAPFASTER}                                                & \multicolumn{1}{c|}{$\text{SBS}_{\text{Acc}}$}        & \multicolumn{1}{c}{$\text{SBS}_{\text{Gap}}$} \\ \cline{2-14} 
Dataset & Task                                                 & \multicolumn{1}{c}{Acc $\uparrow$}& \multicolumn{1}{c|}{Gap $\downarrow$}        & \multicolumn{1}{c}{Acc $\uparrow$}& \multicolumn{1}{l|}{Gap $\downarrow$}     & \multicolumn{1}{c}{Acc $\uparrow$} & \multicolumn{1}{l|}{Gap $\downarrow$}    & \multicolumn{1}{c}{Acc $\uparrow$} & \multicolumn{1}{c|}{Gap $\downarrow$}  & \multicolumn{1}{c}{Acc $\uparrow$} & \multicolumn{1}{c|}{Gap $\downarrow$}    & \multicolumn{1}{c|}{Acc $\uparrow$}                   & \multicolumn{1}{c}{Gap $\downarrow$} \\ \hline
&Score-0.001-CE                             & 0.81                    & 21.07                    & 0.81                    & 19.03                    & 0.84                    & 20.49                    & \textbf{0.91}           & 2.60            & \textbf{0.80}          & \textbf{1.00}            & 0.80                      & 1.00                       \\
&Score-0.001-BCE                            & \textbf{0.85}           & 4.48                     & \textbf{0.83}           & 18.67                    & \textbf{0.85}           & 22.37                    & 0.86                    & 50.94            & \textbf{0.80}          & \textbf{1.00}            & 0.80                      & 1.00                       \\
&Score-0.001-Reg                            & 0.80                    & \textbf{0.89}            & 0.79                    & \textbf{0.87}            & 0.80                    & \textbf{1.00}            & 0.80                    & \textbf{0.94}    & \textbf{0.80}          & \textbf{1.00}            & 0.80                      & 1.00                       \\ \cline{2-14}
&Score-0.1-CE                               & 0.61                    & 16.73                    & 0.61                    & 14.83                    & \textbf{0.69}           & 12.90                    & \textbf{0.61}           & 1.11             & \textbf{0.57}          & \textbf{1.00}            & 0.57                      & 1.00                       \\
&Score-0.1-BCE                              & \textbf{0.65}           & 10.74                    & \textbf{0.62}           & 14.79                    & 0.67                    & 9.20                     & 0.56                    & 2.55             & \textbf{0.57}          & \textbf{1.00}            & 0.57                      & 1.00                       \\
&Score-0.1-Reg                              & 0.57                    & \textbf{0.75}            & 0.57                    & \textbf{0.71}            & 0.57                    & \textbf{1.00}            & 0.57                    & \textbf{1.00}    & \textbf{0.57}          & \textbf{1.00}            & 0.57                      & 1.00                       \\ \cline{2-14}
Standard&Score-1-CE                         & \textbf{0.62}           & 1.82                     & \textbf{0.69}           & 1.95                     & \textbf{0.66}           & 2.09                     & 0.60                    & 3.52             & \textbf{0.62}          & 2.53           & 0.62                      & 1.00                       \\
&Score-1-BCE                                & 0.57                    & 4.42                     & \textbf{0.68}           & 1.89                     & 0.64                    & 1.96                     & \textbf{0.62}           & 2.53             & \textbf{0.62}          & 2.53            & 0.62                      & 1.00                       \\
&Score-1-Reg                                & 0.56                    & \textbf{0.71}            & 0.30                    & \textbf{0.92}            & 0.26                    & \textbf{1.00}            & 0.47                    & \textbf{0.75}    & 0.26          & \textbf{1.00}            & 0.62                      & 1.00                       \\ \cline{2-14}
&Bound-1.1-CE                               & 0.56                    & 2.99                     & 0.65                    & 2.66                     & 0.69                    & 2.81                     & \textbf{0.61}           & 4.24             & \textbf{0.51}          & \textbf{2.53}            & 0.51                      & 1.00                       \\
&Bound-1.2-CE                               & \textbf{0.72}           & 1.64                     & \textbf{0.74}           & 2.07                     & \textbf{0.73}           & 4.59                     & 0.58                    & 2.20             & 0.19          & \textbf{2.53}            & 0.51                      & 1.00                       \\ \hline
& Score-0.001-CE  & \textbf{0.70}           & 1.11                    & \textbf{0.71}            & 1.12                    & \textbf{0.70}            & 0.97                    & \textbf{0.69}            & 1.00                    & 0.69             & \textbf{1.00}          & \textbf{0.69}                     & 1.00                       \\
EECBS & Score-0.001-BCE & 0.69              & 1.05                    & 0.69                     & \textbf{1.00}           & \textbf{0.70}            & \textbf{0.93}           & \textbf{0.69}            & 1.00                    & 0.69             & \textbf{1.00}          & \textbf{0.69}                     & 1.00                       \\
&Score-0.001-Reg & 0.69                     & \textbf{1.00}           & 0.69                     & \textbf{1.00}           & 0.69                     & 1.00                    & \textbf{0.69}            & \textbf{0.93}           & 0.69             & \textbf{1.00}          & \textbf{0.69}                     & 1.00                       \\
\end{tabular}

}
    \caption{The accuracy (Acc) and VBS-SBS gap (Gap) results for different models and different learning tasks in terms of accuracy and VBS-SBS gap. The names of the tasks are shown in the Task column, following the naming described in the experiment setting section. The best results for each metric on the same optimization objectives are marked with bold (Bound tasks do not have bold in Gap metrics because in bound tasks accuracy is normally the primary focus).}
\end{table*}

\begin{table*}[h!]
\centering
\begin{tabular}{@{}c|cc@{}}
                        & VGGNet                             & ViT                                \\ \hline
input size              & 384                                & 384                                \\
optimizer               & Adams                              & Adams                              \\
learning rate            & 3e-6                               & 3e-6                               \\
learning rate search space         & 1e-4, 5e-5, 3e-5, 1e-5, 3e-6, 1e-6 & 1e-4, 5e-5, 3e-5, 1e-5, 1e-6       \\
weight decay            & 3e-2                               & 0                                  \\
optimizer momentum      & $\beta_1,\beta_2 = 0.9, 0.999$     & $\beta_1,\beta_2 = 0.9, 0.999$     \\
batch size              & 64                                 & 64                                 \\
training epochs         & 80                                 & 80                                 \\
learning rate scheduler & Linear                             & Linear                             \\
warmup                  & 5                                  & 0                                  \\
random flipping         & 0.5                                & 0.5                                \\
random rotate           & 0.5                                & 0.5                                \\
random erasing          & 0.5                                & 0.5                      
\end{tabular}
    \caption{Hyperparameter table with search space.}
    \label{tab:experiment_results1}
\end{table*}

\section{Additional Dataset Analysis}

\begin{figure}[h]
    \centering
    \includegraphics[width=0.95\linewidth]{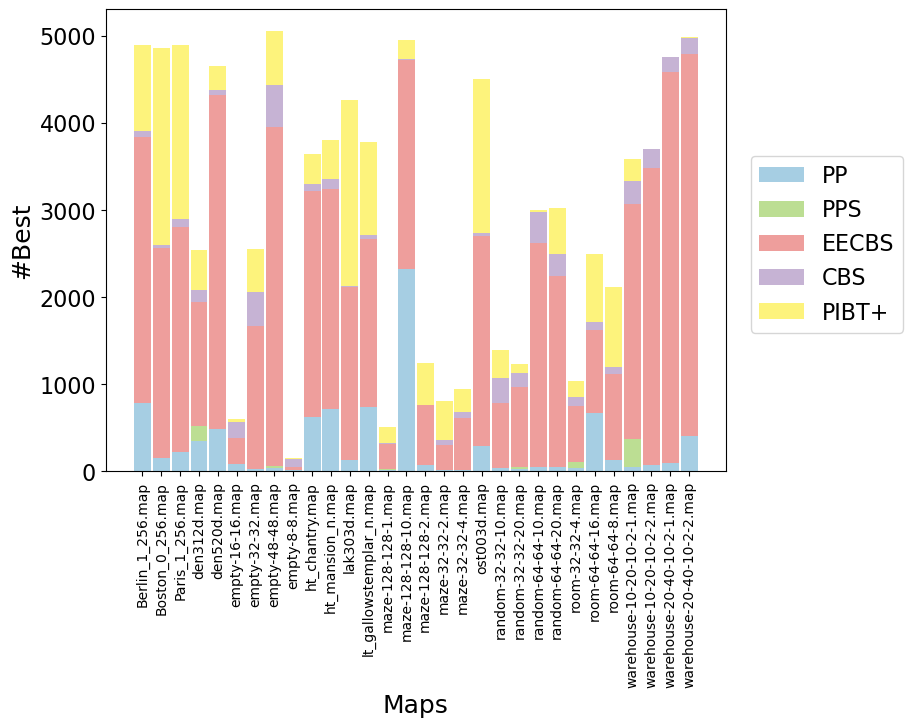}
    \caption{Dataset result for the case of $score=time+1 \times cost$ as score definition.}
    \label{fig:st}
\end{figure}

\begin{figure}[h]
    \centering
    \includegraphics[width=0.95\linewidth]{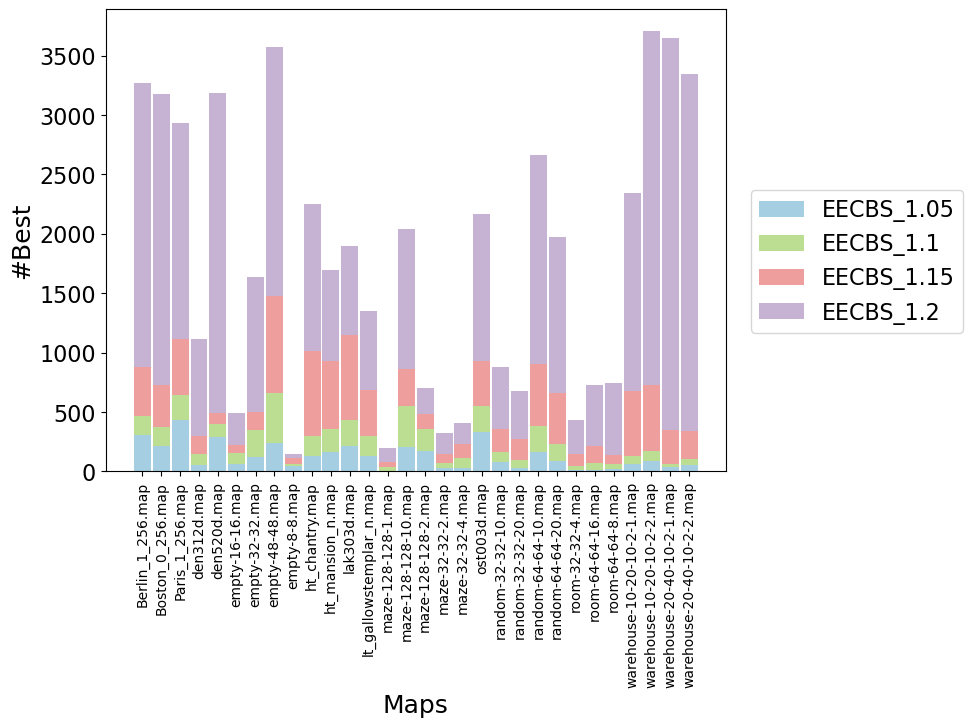}
    \caption{Dataset result for the case of $score=time+0.001 \times cost$ as score definition on hyperparameter selection (EECBS) dataset.}
    \label{fig:ed}
\end{figure}

While we have only shown dataset illustrations of only one objective in the paper, in Fig.~\ref{fig:st} - \ref{fig:ed} we provide the dataset illustrations for all tasks used in the experiment. We can see that when the objective is different, the relative performance of each algorithm is also different. However, there is always one algorithm that wins the single best solver in a quite dominant way, making the dataset very imbalanced.

\begin{figure}[h]
    \centering
    \includegraphics[width=0.95\linewidth]{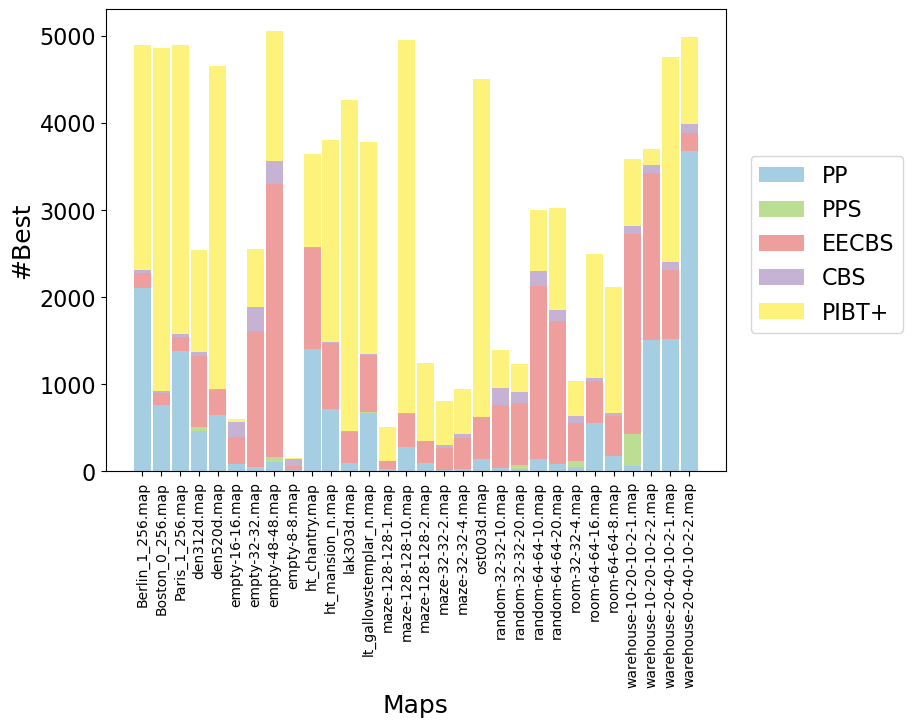}
    \caption{Dataset result for the case of $score=time+0.1 \times cost$ as score definition.}
\end{figure}

\section{Parameter Details}

\begin{table*}[h]
\centering
\begin{tabular}{@{}c|cc@{}}
                        & MAPFAST                            & ResNet                             \\ \hline
input size              & 384                                & 384                                \\
optimizer               & Adams                              & Adams                              \\
learning rate           & 1e-5                               & 1e-5                               \\
learning rate search space         & 1e-4, 5e-5, 3e-5, 1e-5, 3e-6, 1e-6 & 1e-4, 5e-5, 3e-5, 1e-5, 3e-6, 1e-6 \\
weight decay            & 0                                  & 3e-2                               \\
optimizer momentum      & $\beta_1,\beta_2 = 0.9, 0.999$     & $\beta_1,\beta_2 = 0.9, 0.999$     \\
batch size              & 64                                 & 64                                 \\
training epochs         & 80                                 & 80                                 \\
learning rate scheduler & Linear                             & Linear                             \\
warmup                  & 0                                  & 5                                  \\
random flipping         & 0.5                                & 0.5                                \\
random rotate           & 0.5                                & 0.5                                \\
random erasing          & 0.5                                & 0.5                 
\end{tabular}
    \caption{Continuing hyperparameter table with search space.}
    \label{tab:experiment_results2}
\end{table*}

Here we provide our detailed hyper parameter table with our search space in Table.~\ref{tab:experiment_results1} and Table.~\ref{tab:experiment_results2}. Due to limited computing resources, for MAPFASTER, we only use the parameter proposed by the original paper.

\begin{figure}[t]
    \centering
    \includegraphics[width=0.5\linewidth]{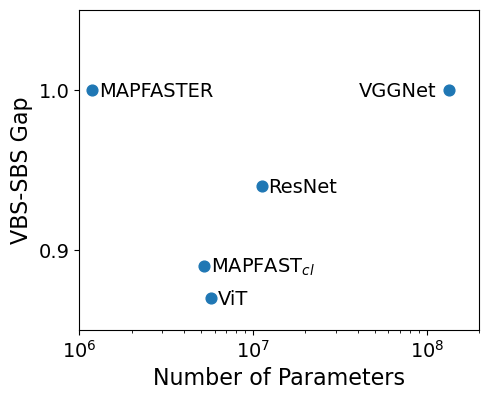}
    \caption{Comparing the gap for different neural networks on the Score-0.001-Reg task based on parameter count.}
    \label{fig:result_n_params}
\end{figure}

\section{Discussion on Choosing a Neural Network for Algorithm Selection}
While we have found that ViT is the neural network that wins the most time, 
it is also interesting to see how results change according to the number of parameters it has, which affect both the training time and the inference time. We conclude our findings in Fig.~\ref{fig:result_n_params}. We find that while we have used the smallest variants of each group of neural networks, ViT is the one that not only has a small number of parameters but also has a good performance. Other modern models can also be successful in tasks while small in the number of parameters. On the other hand, when we use the ViT-Large model instead of ViT-Base model on the group of tasks with w=0.001, we found the performance is not improving. This is because our dataset is relatively too small to stop a large model from overfitting the dataset.

\end{document}